\documentclass{aa}
\usepackage{graphicx}
\usepackage{natbib}  
\usepackage{txfonts}
\def\msun{\ensuremath{{M}_{\odot}}}

\def\micron{$\mu$m~}

\begin{document}
   \title{An empirical mass-loss law for Population II giants from the Spitzer-IRAC survey of Galactic globular
   clusters
\thanks{This work is based on observations made with the Spitzer Space Telescope, 
which is operated by the Jet Propulsion Laboratory, 
California Institute of Technology under a contract with NASA.
Support for this work was provided by NASA through 
an award issued by JPL/Caltech.}}


   \author{L. Origlia\inst{1}
	  \and F.R. Ferraro\inst{2}
          \and S. Fabbri\inst{2} 
	  \and F. Fusi Pecci\inst{1} 
	  \and E. Dalessandro\inst{2}
          \and R.M. Rich\inst{3} 
	  \and E. Valenti\inst{4} 
          }

\institute{
             INAF - Osservatorio Astronomico di Bologna,
             Via Ranzani 1, I-40127 Bologna, Italy
	     \email{livia.origlia@oabo.inaf.it}
         \and
             University of Bologna, Physics \& Astronomy Dept.,
             Viale Berti Pichat 6-2, I-40127 Bologna, Italy
         \and
	  Department of Physics and Astronomy, University of California at Los Angeles, Los Angeles, CA 90095-1547, US
         \and
	 ESO - European Southern Observatory, Karl-Schwarzschild Str. 2, D-85748 Garching bei M\"unchen, Germany
             }

\authorrunning{Origlia et al.}
\titlerunning{Mass loss in Population II giants}

   \date{Received .... ; accepted ...}

 
  \abstract
  {}
   {The main aim of the present work is to derive an empirical mass-loss (ML) law for Population II stars in first and second ascent red giant branches.}
   {We used the Spitzer InfraRed Array Camera (IRAC) photometry obtained in the 3.6--8\,\micron range of a carefully 
chosen sample of 15 Galactic globular clusters spanning the entire metallicity 
range and sampling the vast zoology of  horizontal branch (HB) morphologies.
We complemented the IRAC photometry with near-infrared data to build suitable
color-magnitude and color-color
diagrams and identify mass-losing giant stars.}
{We find that  
while the majority of stars show colors typical of cool giants,
some stars show an excess of mid-infrared light that is larger than expected from
their photospheric emission and that is 
plausibly due to dust formation in mass flowing from them. 
For these stars, we
estimate dust and total (gas + dust) ML rates and timescales. We finally calibrate an
empirical ML law for Population II red and asymptotic giant branch stars with varying metallicity.
We find that at a given red giant branch luminosity only a fraction of the stars are
losing mass. From this, we conclude that ML is episodic and is active 
only a fraction of the time, which we define as the duty cycle.  
The fraction of mass-losing stars increases by increasing the stellar luminosity and  
metallicity.
The ML rate, as estimated from reasonable assumptions for the gas-to-dust ratio 
and expansion velocity, depends on metallicity and slowly increases with decreasing
metallicity. In contrast, the duty cycle increases with increasing
metallicity, with the net result that total ML increases
moderately with increasing metallicity, about 0.1\,\msun\ every dex in
[Fe/H].  For Population II asymptotic giant branch stars, we
estimate a total ML of $\le 0.1$\,\msun, nearly constant with varying
metallicity.  
}
   {}

   \maketitle
%

\section{Introduction}

Mass loss (ML) affects all stages of stellar evolution and its parametrization 
remains a vexing problem in any modeling,
since satisfactory empirical determinations
as well as a comprehensive physical description of the involved
processes are still lacking.
This is especially true for Population II red giant branch (RGB) and
asymptotic giant branch (AGB) stars.

The astrophysical impact of ML in Population II giants is
huge and affects not only stellar evolution modeling, but also
related subjects, like, for example, the UV excess in ellipticals or
the interaction between the cool intracluster medium and hot halo gas.
There is a great deal of indirect, but quantitative evidence for ML
during the RGB evolution, namely the horizontal branch (HB) morphology and the 2nd parameter problem, the pulsational
properties of RR Lyrae, the absence of AGB stars significantly
brighter than the RGB tip, and the masses of white dwarfs (WDs) in
Galactic globular clusters (GCs) \citep[see,
e.g.,][]{roo73,ffp75,ffp76,ren77,ffp93,fer98,cru96,han05,kal07,cat09}.  On the
contrary, there is no empirical ML law directly calibrated on
Population II giants with varying metallicity and only a few estimates
of ML for giants on the brightest portion of the RGB and AGB exist.  
As a consequence, ML timescales, driving
mechanisms, dependence on stellar parameters, and metallicity are still
open issues. There is little theoretical or observational
guidance on how to incorporate ML into models.

With no better recipe, models of stellar evolution incorporate ML
by using analytical ML formulae calibrated on bright Population I
giants. The first and most used of these is the \citet{rei75a,rei75b}
formula, extrapolated toward lower luminosity and introducing a free
parameter $\eta$ (typically equal to 0.3) to account for a somewhat less
efficient ML along the RGB.  A few other formulae, which are variants
of the Reimers formula, have been proposed in the subsequent years
\citep[see, e.g.,][]{mul78,gol79,jud91}.  More recently, \citet{cat00}
revised these formulae by using a somewhat larger database of stars
than in previous studies, but still amounting to 20--30 giants only,
the majority being AGB stars.  \citet{sc05} propose a new
semi-empirical formula that explicitly includes a dependence from all
the stellar parameters.  Further advances clearly require empirical
estimates of ML rates in low-mass giants along the entire RGB and AGB
extension.
                   
There are two major diagnostics of ML in giant stars:
the detection of outflow motions in the outer regions of the stellar
atmosphere or the detection of circumstellar (CS) envelopes at much larger
distances from the star.

After the pioneering work by \citet{rei75a,rei75b}, 
the systematic investigation of chromospheric lines in giants stars with possible
emission wings started in the 1980s.  \citet{gra83,cac83,gra84} measured
H$\alpha$ emission in old, bright giants near the RGB tip, members of
Galactic globular and open clusters.  They found H$\alpha$ emission in
a significant fraction of them and, by using the simple recombination
model by \citet{cohen76}, they estimated average $ dM/dt \approx
10^{-8} M_{\odot}\,{\rm yr^{-1}}$ ML rates.  However,
\citet{dup84} and \citet{dup86} argued that the H$\alpha$ wings could naturally
arise in a static stellar chromospheres.  Other authors
\citep[e.g.,][]{peterson81,peterson82,dup92,dup94,lyons96,smith04,cac04,mau06,vie11}
investigated the possible presence of profile asymmetries and
coreshifts in a large number of chromospheric lines, by means of high
resolution spectroscopy over a wide spectral range, from UV (MgII h,k
$\lambda$2800 \AA) to optical (CaII K, NaI D, H$\alpha$) and IR (HeI
$\lambda$10830.3 \AA).  These line asymmetries and coreshifts can be
accounted for only by an active chromosphere and/or mass outflow,
with typical velocity fields of 10--20 km/s.  The difficulty of
converting the chromospheric line diagnostics into ML rates is
certainly related to modeling uncertainties, for example because of the lack
of any detailed knowledge of the structure and excitation mechanism of
the wind region.  However, it is also clear that the outflow region
traced by the chromospheric lines is still too close to the star, to
sample the bulk of the mass lost, likely accumulated at larger
distances.  Hence, the chromospheric line method seems more effective
in tracing the region of wind formation and acceleration, rather than
most of the outflow.  Finally, it must be recalled that even with
8m-class telescopes it is at best expensive and often impossible to
obtain high-resolution, high S/N spectra of Population II giants along
the entire RGB extension.

A CS envelope around a cool giant can be detected by measuring IR dust
emission, linear polarization, microwave CO emission and radio OH
masers.  However, CS envelopes of low-mass giants have intrinsically
low surface brightness.  Far IR and radio receivers have neither
sufficient spatial resolution nor sensitivity to study Population II
CS envelopes in dense stellar fields.  Linear polarization,
intrinsically well below 1\%, is also hardly measurable.  Hence, array
photometry in the 3--20\,\micron region remains the most effective
way to detect Population II CS envelopes.  Mid-IR observations
have the advantage of sampling an outflowing gas fairly far from the
star (typically, tens/hundreds stellar radii).  Such gas left the star
a few decades previously, hence the inferred ML rate is also smoothed
over such a timescale.  In the late 1980s, the first measurements of dust
excess in Galactic GC giants by means of mid-IR
photometry from the ground \citep{frog88} and with IRAS
\citep{gil88,ori96} became available, although the spatial resolution
of these detectors was insufficient to properly resolve most of the
stars.  A decade later, the Infrared Space Observatory (ISO) satellite allowed new observations,
but was still limited in spatial resolution and sensitivity.  A few bright
AGB stars in 47~Tuc have been measured by \citet{ram01}, finding dust
excess in two objects only.  Our group performed a deep survey with
the ISO Infrared Camera ISOCAM of six  massive GCs \citep{ori02}, namely 47~Tuc,
NGC~362, $\omega$ Cen, NGC~6388, M~15 and M~54, in the 10\,\micron
window.  From a combined physical and statistical analysis, our ISOCAM
study provided ML rates and frequency for some giants near the tip
\citep[see also][]{ori07}.  However, the small sample of observed
giants and the limited capabilities of ISOCAM
allowed us to reach only weak conclusions on the ML dependence on
luminosity, metallicity, and HB morphology.

The advent of Spitzer with its mid-InfraRed Array Camera (IRAC) has opened a new
window in the study of CS envelopes around Population II giants.
Indeed, the IRAC bands between 3.6 and 8\,\micron are effective in
detecting warm dust with spatial resolution good enough to resolve a
large fraction of the GC giants.  By using the $(3.6-8)$ Spitzer-IRAC
color as a diagnostic, dust excess has been detected around some of the
brightest giants in $\omega$ Cen, M~15, NGC~362 and 47~Tuc
\citep{boy06,boy08,boy09,boy10}.

In Cycle 2 (program ID \#20298), our group was granted 26\,hr of
Spitzer-IRAC observing time to map 17 Galactic GCs  down to the HB
level.  
We combined Spitzer-IRAC photometry with high-resolution near-IR
photometry from the ground and used (K-IRAC) colors as diagnostics of
possible circumstellar dust excess.  Results for 47~Tuc have been
published in \citet{ori07,ori10}, while those for the complex stellar system $\omega$ Cen will 
be presented in a forthcoming paper.  

Here we present the photometric analysis for the
remaining 15 GCs in our sample,  we discuss ML rates
and duty cycles in Population~II giants and we derive 
an empirical
law of ML for Population II giants with varying metallicity.

\section{Observations, data reduction, and photometric analysis}
\label{obs}

\begin{figure*}[htbp]
 \centering
  \includegraphics[width=13.5cm]{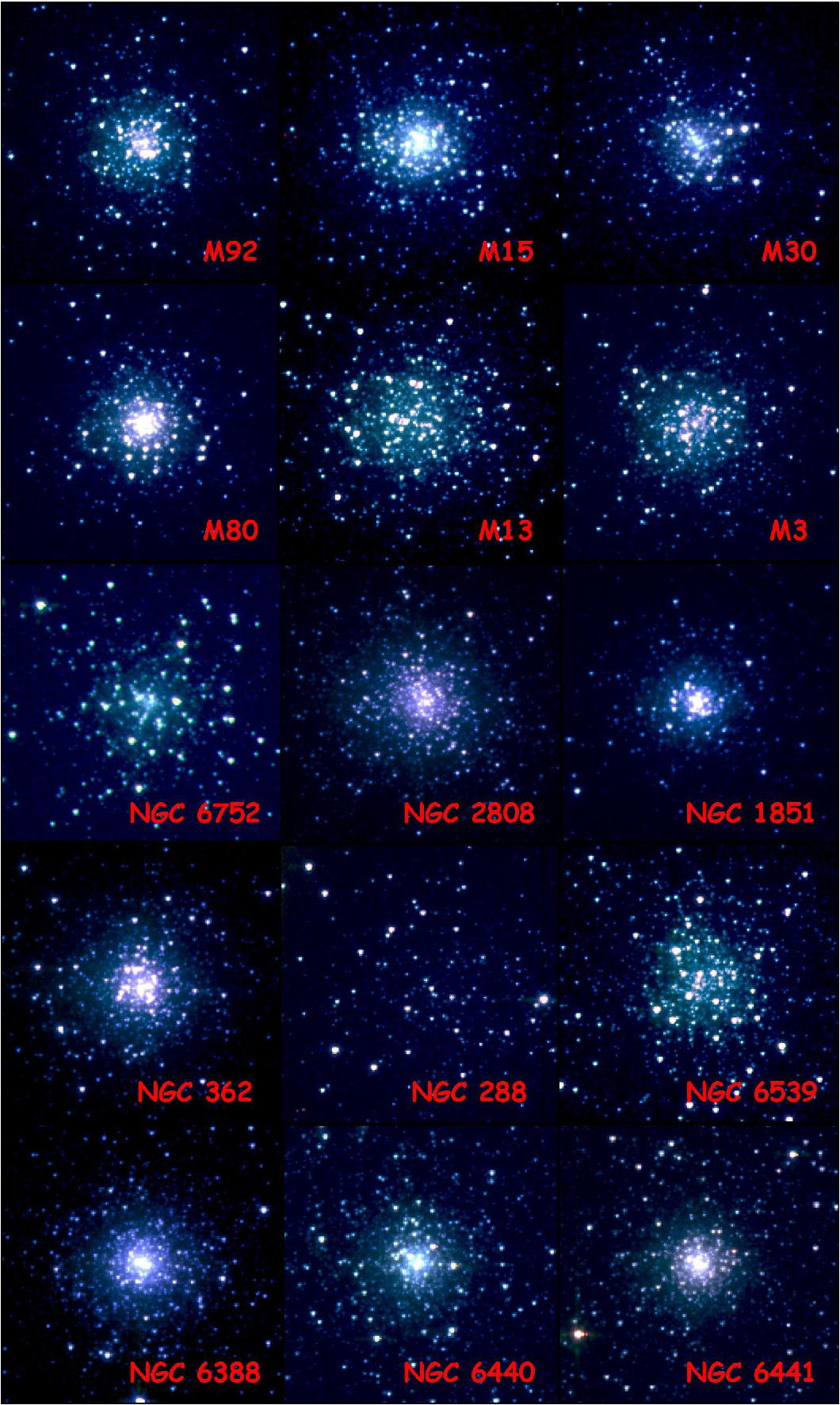}
\caption{Spitzer-IRAC 
three color (3.6$\mu$m (blue), 6$\mu$m (green), 8$\mu$m (red)) images of the 
15 GCs in our sample.
\label{imaspitzer}}
\end{figure*}

The 15 GCs presented in this paper are listed in Table~1.
They span metallicities from about 1/100 to about solar; at a
given metallicity, clusters with different HB morphologies were chosen.

\begin{table*}[htbp]
\begin{center}
\caption{Observed sample of GCs and main parameters.}
\begin{tabular}{lllllll}
\hline
Cluster &  [Fe/H]$^a$ &$(m-M)_0^a$ & $E(B-V)^a$ & [M/H]$^b$& HB$^c$ & $t_{\rm exp}$$^d$\\
\hline
NGC~288                 &$-1.07$ & 14.73 & 0.03     &$-1.00$ & E&1.10hr  \\
NGC~362                 &$-1.15$ & 14.68 & 0.05     &$-0.94$ & N&1.10hr  \\
NGC~1851                &$-1.08$ & 15.46 & 0.02     &$-0.87$ & E&2.05hr  \\
NGC~2808                &$-1.15$ & 14.90 & 0.23     &$-0.94$ & E&2.05hr  \\
NGC~5272 (M3)           &$-1.34$ & 15.03 & 0.01     &$-1.13$ & N&2.05hr  \\
NGC~6093 (M80)          &$-1.41$ & 14.96 & 0.18     &$-1.20$ & E&2.05hr  \\
NGC~6205 (M13)          &$-1.39$ & 14.43 & 0.02     &$-1.18$ & E&1.10hr  \\
NGC~6341 (M92)          &$-2.16$ & 14.78 & 0.02     &$-1.95$ & N&1.10hr  \\
NGC~6388                &$-0.61$ & 15.30 & 0.44     &$-0.40$ & E&2.05hr  \\
NGC~6539                &$-0.66^e$ & 14.62 & 1.08   &$-0.45$ & N&1.10hr  \\
NGC~6440                &$-0.56^e$ & 14.48 & 1.15   &$-0.35$ & N&1.10hr  \\
NGC~6441                &$-0.50^e$ & 15.26 & 0.52   &$-0.29$ & E&2.05hr  \\
NGC~6752                &$-1.42$ & 13.18 & 0.04     &$-1.21$ & E&0.90hr  \\
NGC~7078  (M15)         &$-2.12$ & 15.15 & 0.09     &$-1.91$ & E&2.05hr  \\
NGC~7099  (M~30)        &$-1.91$ & 14.71 & 0.03     &$-1.00$ & N&1.10hr  \\
\hline
\end{tabular}
\end{center}
\label{clusters}
($a$)~Cluster metallicity, reddening, and distance modulus from \citet{fer99a,fer00} and
\citet{val07}.\\
($b$)~Cluster global metallicity computed using the formula 
$({\rm [M/H]}={\rm [Fe/H]}+\log_{10} (0.638\times f_{\alpha}+0.362)$, 
by assuming an overall [$\alpha/{\rm Fe}]=+0.3$ enhancement 
(i.e., $f_{\alpha}=2$) \citep[see][]{sal93,fer99a}.\\
($c$)~HB morphology: N=normal (red and blue if metal-poor clumps), E=extended (blue tails).\\                            
($d$)~Total exposure time for our IRAC mapping, according to the
AOR duration estimated with SPOT--v11.07.\\
($e$)~Cluster metallicity from high-resolution IR spectroscopy by \citet{ori05} and
\citet{ori08}.\\
\end{table*}

\begin{itemize}
\item {\bf M~92, M~15, M~30} --
These are the most metal-poor clusters in our sample, with M~15 and, to a lesser extent, M~30, 
showing very blue HB tails compared to M~92. 

\item{\bf M80, M13, M3} --
This is the popular cluster triplet at intermediate metallicity 
with different HB morphologies, including red clumps, gaps, and blue tails
\citep[see, e.g.,][]{fer97,fer98,dal13a}.

\item{\bf NGC~6752, NGC~288, NGC~2808, NGC~362, NGC~1851} --
This group of clusters at intermediate metallicity also shows very different  
HB morphologies, with purely blue (NGC~6752, NGC~288), purely red (NGC~362), and multimodal 
distributions with blue tails (NGC~2808 and NGC~1851). 

\item{\bf NGC~6440, NGC~6388, NGC~6441, NGC~6539} --
These are the most metal-rich clusters in our sample. 
While NGC~6440 and NGC~6539 have a red HB, typical of their high metallicity, 
NGC~6441 and NGC~6388 have a multimodal HB distribution, with a well populated 
red clump and a peculiar blue HB tail 
\citep{ric97,pri01,pri02,pri03,bus07,dal08}.
\end{itemize}

For sake of clarity, we grouped them in two classes according to their HB morphology, 
namely N=normal if they have red and blue (if metal-poor) clumps  and E=extended if they 
have blue tails (see Table~1). 

We observed these GCs between September 2005 and July 2006, using
the mid-IR camera IRAC, onboard the Spitzer Space Telescope.
A frame time of 12 sec and a {\em 25--30 cycling position} dithering
pattern with the {\em small} scale factor, repeated a few times for on-source 
integration times between 1000\,s and 2700\,s, and total observing
time between 1.1hr and 2.05\,hr, allowed us to cover a $5' \times 5'$
field of view in all the four IRAC channels, at 3.6, 4.5, 5.8, and 8.0
\,\micron.  
Figure~\ref{imaspitzer} shows the three color (3.6$\mu$m (blue), 6$\mu$m (green), 8$\mu$m (red)) 
mosaicked images for the 15 observed GCs.

The [Post Basic Calibrated Data] mosaic frames from the Spitzer Pipeline
(Software Version: S13.2.0), with signal per pixel in unit of MJy~sr$^{-1}$,
have been photometrically reduced with
ROMAFOT \citep{buo83}, a software package optimized for point spread
function (PSF) fitting in crowded and undersampled stellar fields. 

We used a constant PSF over the entire IRAC field of view.
Indeed, we verified that in all of our IRAC images the PSF is constant within $\pm$2\%. 
Such a small (if any) variation of the PSF has a negligible impact on the computed 
magnitudes (a few hundredths of a magnitude  at most), 
well within the overall error budget of our Spitzer photometry (see Appendix A3). 

We converted the instrumental magnitudes of each star 
into the Vega magnitude system by 
using the zero-magnitude flux densities of \citet{rea05}.

We obtained complementary ground-based near-IR photometry of the central region at
higher, sub-arcsec spatial resolution at ESO, La
Silla (Chile) and at the TNG for the southern and northern clusters,
respectively.  A detailed description of the observations can be found
in \citet{val04a,val04b,val07}. 
Both instrumental
magnitudes and star positions were placed into the Two Micron All
Sky Survey (2MASS) system.  Then we supplemented these photometric catalogs 
with 2MASS data in the most external regions. 
We cross-correlated the
Spitzer-IRAC catalogs with the near-IR ones 
and we constructed a final catalog for each cluster, including stars with both near- and mid-IR photometry. 
The fact that the Spitzer-IRAC 
catalogues are combined with high-resolution near-IR stellar photometric catalogues,
suitably cleaned by spurious detections and background
sources/galaxies, also has the advantage of automatically excluding
contamination by non stellar objects and/or IRAC instrumental artifacts from the 
final Spitzer-IRAC catalogues. 

We corrected the IR magnitudes for extinction using the
$E(B-V)$ values reported in Table~1 and the \citet{rie85}
and \citet{ind05} interstellar extinction laws.

\section{Color-magnitude and color-color diagrams}
\label{diag}

The combination of near-IR and Spitzer-IRAC photometries allows for many 
effective diagnostic planes to search for stars with color excess, 
thus tracing the presence of dusty CS envelopes.

The 1--2.5\,\micron spectral range is especially suitable to measure
the photospheric emission of cool stars.  
We computed suitable transformations between near-IR colors and bolometric corrections
and effective temperatures as obtained from the Kurucz's model spectra
convolved with the 2MASS broad-band filters.
Very similar transformations were obtained by \citet{mon98} using a large database of 
observed colors for GC giants and suitable grids of different model atmospheres.
In this respect, we note that old GC stars have age, distance, and reddening 
determined with great accuracy and consequently their mass and 
photospheric parameters can be constrained by using more robust 
calibrations than a mere spectral energy distribution fitting, as used for field stars. 
The (J-K) color is especially effective in sensing the temperature of low-mass RGB and AGB stars 
and these temperature estimates are in very good agreement (normally well within 100 K) 
with those obtained from other colors like the (V-K) or the (V-I). 
Bolometric corrections in the K-band are very effective for estimating the bolometric magnitude. 
Hence, for each star detected in our Spitzer-IRAC survey we
derived the bolometric magnitude and 
temperature from its dereddened $(J-K)_0$ color and $M_K$ absolute magnitude, by  
using the distance moduli by \citet{fer99a,fer00} and \citet{val07}.
The use of bolometric magnitudes allows for a direct 
comparison of color-magnitude diagrams (CMDs) from different GCs and also 
with model predictions.  

In very cool, luminous
giants the 3--5\,\micron spectral range is still largely dominated by
the photospheric emission. However, CS dusty envelopes can also
radiate in this spectral range. Indeed, in relatively warm and low
luminosity giants, like low-mass RGB stars, the fractional
contribution of warm, optically thin dust emission from a CS envelope
is not negligible in the 3--5\,\micron spectral range.
Hence, as detailed in \citet{ori10},
a combination of near- and mid-IR colors like
$(K-5.8)$ and $(K-8)$, is more effective in tracing the possible
presence of small amounts of warm dust around low-mass RGB stars than
the pure Spitzer-IRAC $(3.6-8)$ color, which is mostly sensitive 
to detect relatively large amount of cold dust around the coolest (hence the most
luminous) giants.
Since the 8\,\micron IRAC band is the most sensitive to warm
dust and the least contaminated by photospheric emission, we use the
$(K-8)_0$ color as a primary diagnostic to select stars with possible
dust excess. 

\begin{figure*}[htbp]
\begin{center}
\includegraphics[width=\hsize]{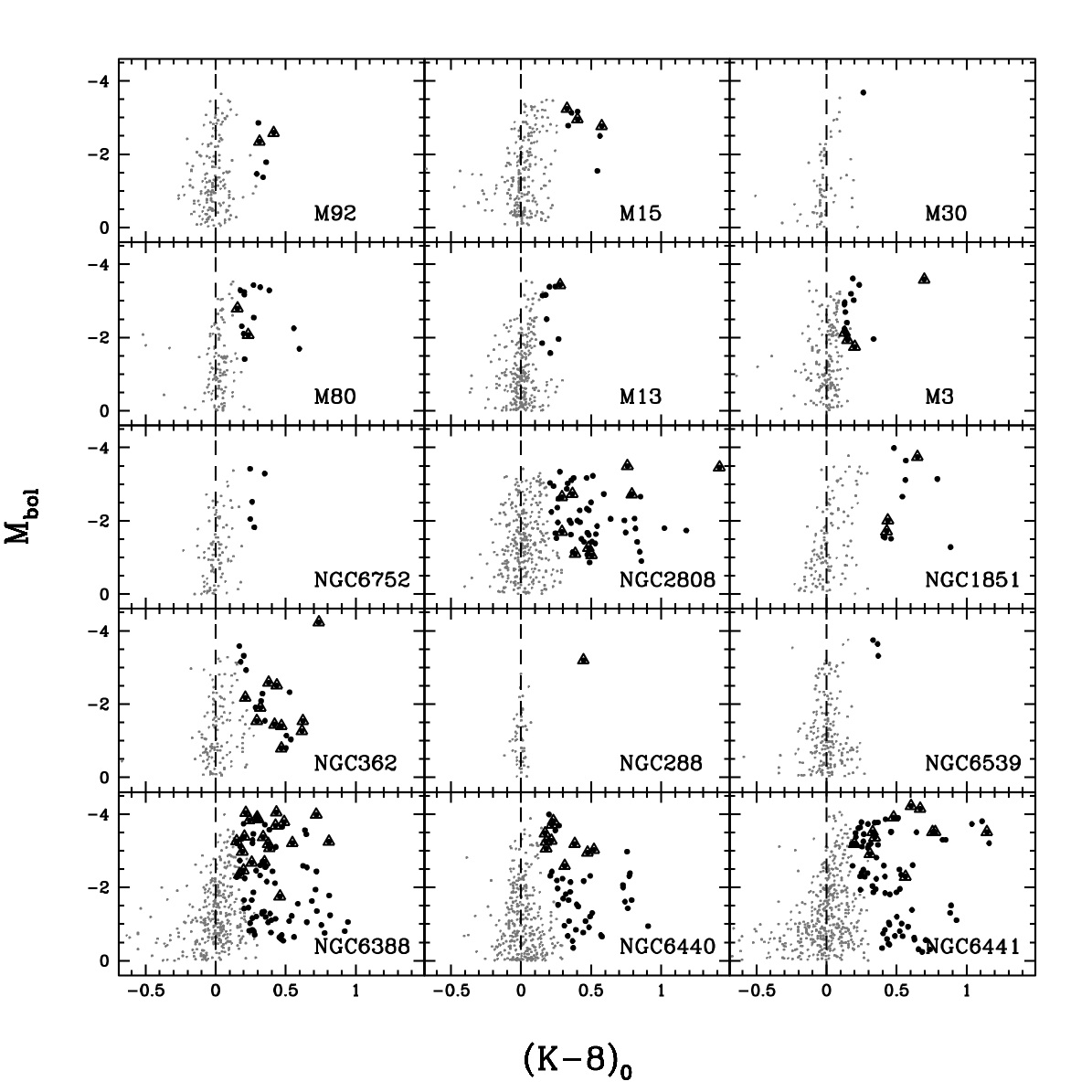}
\end{center}
\caption{M$_{bol}$, (K-8)$_0$ CMDs of the 15 GCs in our sample. 
The mean ridge lines centered at (K-8)$_0$=0 are also plotted (dashed lines).
Stars with color excess are marked with black dots, candidate dusty AGB stars
as triangles.
\label{cmd}}
\end{figure*}

We first constructed suitable $M_{\rm bol},~(K-8)_0$
CMDs 
for the 15 GCs in our sample, reported in 
Figure~\ref{cmd}.
The results are fully consistent with a vertical, ridge line centered at color (K-8)$_0$$\approx$0. 
Indeed, small zero point shifts, if any 
(on average 0.00$\pm$0.04), with respect to a nominal value of zero were obtained. 
Although very small, we apply these offsets to the measured (K-8)$_0$ color of all the sampled stars in each 
GC, in order to consistently get the mean ridge line centered at (K-8)$_0$=0.  
This observational evidence is also 
predicted by models.
In stars with no extra-flux, 
the 8 micron filter measures only the photospheric emission of the giants, 
and for example, theoretical model atmospheres from the Kurucz's
database in the temperature range  $T_{\rm eff}=3500$--5000~K predict $(K-IRAC)_0$ colors 
$\approx 0.0 \pm 0.1$ along the entire RGB range sampled by our
survey, which is down to M$_{bol}$$\approx$0.0 mag.

\begin{figure*}[htbp]
\begin{center}
\includegraphics[width=\hsize]{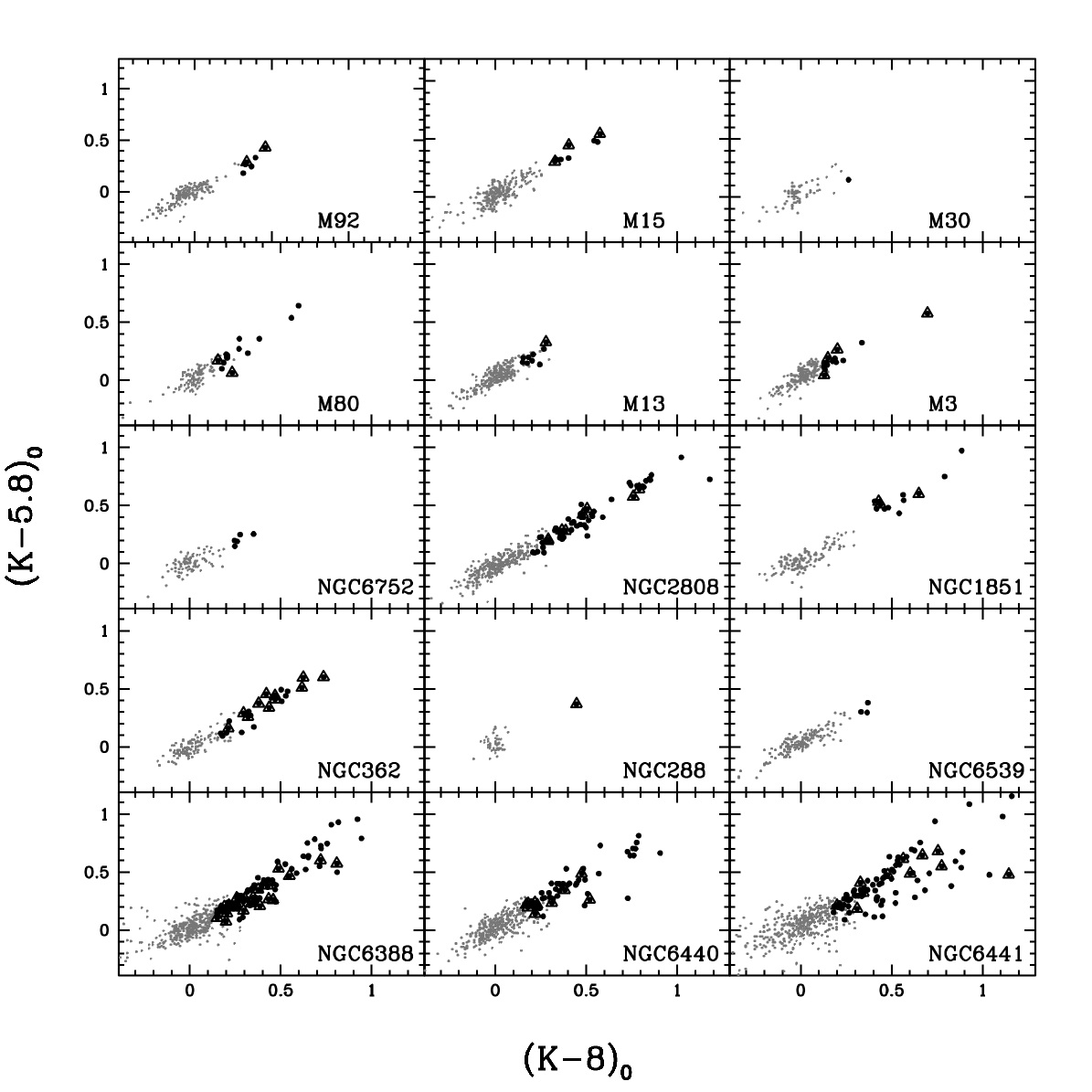}
\end{center}
\caption{$(K-5.8)_0$, $(K-8)_0$ CCDs of the 15 GCs in our sample. 
RGB stars with color excess are marked with black dots, while candidate dusty AGB stars
with filled triangles.
\label{ccd}}
\end{figure*}

Once we established the mean ridge line in each CMD,
we computed the color standard deviation ($\sigma$) for the stars on the blue side of the line 
(i.e., those with sure photospheric emission,  only) in
different magnitude bins.   
Based on this definition, stars on the red side are flagged as ``dusty'' if they show a $(K-8)_0$ 
color excess $\ge +3\sigma$ from the
mean ridge line. We note that these objects are the brightest at the Spitzer wavelengths and therefore, 
they have on average smaller photometric errors (see Appendix). These stars are also normally the reddest in the
other IRAC bands, so as a secondary diagnostic tool we also use the
$(K-5.8)_0,~(K-8)_0$ color-color diagram (CCD) to confirm selection of dusty
star candidates. 
As for the (K-8)$_0$ color, we also measured very small zero point offsets (on average -0.02$\pm$0.04) 
for the (K-5.8)$_0$ color with respect to a nominal value of (K-5.8)$_0$=0.
Figure~\ref{ccd} shows the $(K-5.8)_0,~(K-8)_0$ CCDs for the
15 observed clusters.  

In both the CMDs and CCDs, we marked our final candidate RGB and AGB dusty stars
with filled symbols.
The method used to separate RGB from AGB stars is discussed in Sect.~\ref{fnum}.

\subsection{Mass-losing stars: previous identifications}

The NGC~362, NGC~6388, and M~15 clusters were previously observed with ISOCAM by
\citet{ori02}.  

Three sources in NGC~362 were identified as having IR
excess in our ISOCAM survey. 
\citet{mcd07} obtained VLT/UVES spectra for these three sources, which they designated 
as x01, x02, and x03. All three sources show Spitzer excess, but 
we only confirmed x03 as a genuine dusty
giant, while x01 and x02 turn out to be complex
blends of a few stars with moderate (if any) color excess.  

Six sources in NGC~6388 were identified as having IR excess in
our ISOCAM survey.  All of these sources have also been detected by
Spitzer and confirmed as dusty giants.  

Two sources in M~15 were
identified as having IR excess in ISOCAM survey.  Both sources
have been detected by Spitzer and confirmed as dusty stars.  One of
the ISO sources was also flagged as dusty by \citet{boy06} (their
IR13).

Spitzer observations of M~15 and NGC~362 were analyzed by
\citet{boy06} and \citet{boy09}, respectively.  

\citet{boy06} found 23
dusty IR sources in M~15, but they suggested that these giants are
mostly AGB and post-AGB stars.  Twenty sources are also present in our
catalog, while one (identified as IR4 by \citet{boy06}) is out
of our field of view, and two (identified as IR3a and IR3b by
\citet{boy06}) are too faint (both have $K>15$).  We classify only
their IR13 source as a dusty RGB star, and we also detect the
planetary nebula K648.  

\citet{boy09} find ten candidate mass-losing
stars in NGC~362: they classify s02, s05, s06, s07, s08 as strong mass-loss 
candidates because are very bright and have strong excess; s01,
s03, s04, s09, s10 are identified as moderate ML candidates.
Their s02 and s10 sources are out of our field of view. Sources s05
and s08 appear to be one component with Spitzer, but our near-IR
images reveal that they have two and three components, respectively
(\citet{mcd07} also noted the multiple components of s05). Hence, we
identify s05 and s08 as blends.  Among the remaining six sources we
only classified s01, s06, and s07 as dusty RGB stars.

\section{Star counts, frequency of dusty stars, and ML timescales}
\label{fnum}

\begin{table*}[htbp]
\begin{center}
\caption{Star counts from the photometric Spitzer-IRAC survey.}
\label{phot}
\begin{tabular}{lccccc}
\hline
Cluster &  dusty & dusty &\multicolumn{3}{c}{Fractional numbers$^k$}\\
        &  RGB & AGB+LPV$^a$& $M_{\rm bol}(RGB)\le-1.5$ & -1.5$<$$M_{\rm bol}(RGB)\le -0.6$ &$M_{\rm bol}(AGB)\le-1.5$\\
\hline
NGC~362                 & 12 & 11$^b$  &0.16  & 0.04 &0.40  \\
NGC~1851                &  9 &  3$^c$  &0.13  & 0.02 &0.26 \\
NGC~2808                & 48 &  9$^d$  &0.22  & 0.06 &0.17 \\
NGC~5272 (M~3)          & 11 &  4$^e$  &0.14  & 0    &0.18  \\
NGC~6093 (M~80)         & 12 &  2$^f$  &0.18  & 0.02 &0.10 \\
NGC~6205 (M~13)         &  8 &  1$^e$  &0.09  & 0    &0.12    \\
NGC~6341 (M~92)         &  4 &  2$^g$  &0.03  & 0.02 &0.18\\
NGC~6388                & 69 & 21$^h$  &0.19  & 0.10 &0.27  \\
NGC~6440                & 40 & 10$^b$  &0.24  & 0.10 &0.24  \\
NGC~6441                & 75 & 12$^i$  &0.24  & 0.06 &0.54  \\
NGC~6752                &  5 &  0$^j$  &0.15  & 0    &0.00 \\
NGC~7078 (~M15)         & .5 &  3$^g$  &0.04  & 0    &0.16    \\
\hline
\end{tabular}
\end{center}
($a$)~LPVs from \citet{cle01}.\\
($b$)~From \citet{pio02}, supplemented with WFI photometry (Dalessandro et al. 2013b) in the external region.\\
($c$)~From Lanzoni, private communication, 
supplemented with WFI photometry in the external region.\\
($d$)~From \citet{dal11}.\\
($e$)~From \citet{fer97}.\\
($f$)~From \citet{fer99b}, supplemented with WFI photometry in the external region.\\
($g$)~From Beccari, private communication.\\
($h$)~From \citet{dal08}.\\
($i$)~From Valenti, private communication.\\
($j$)~From \citet{sab04}.\\
($k$)~Fractional number of dusty RGB and AGB stars corrected for incompleteness and field contamination 
(see Section~\ref{fnum} for more details).
\end{table*}

\begin{figure*}[]
\begin{center}
\includegraphics[width=14cm]{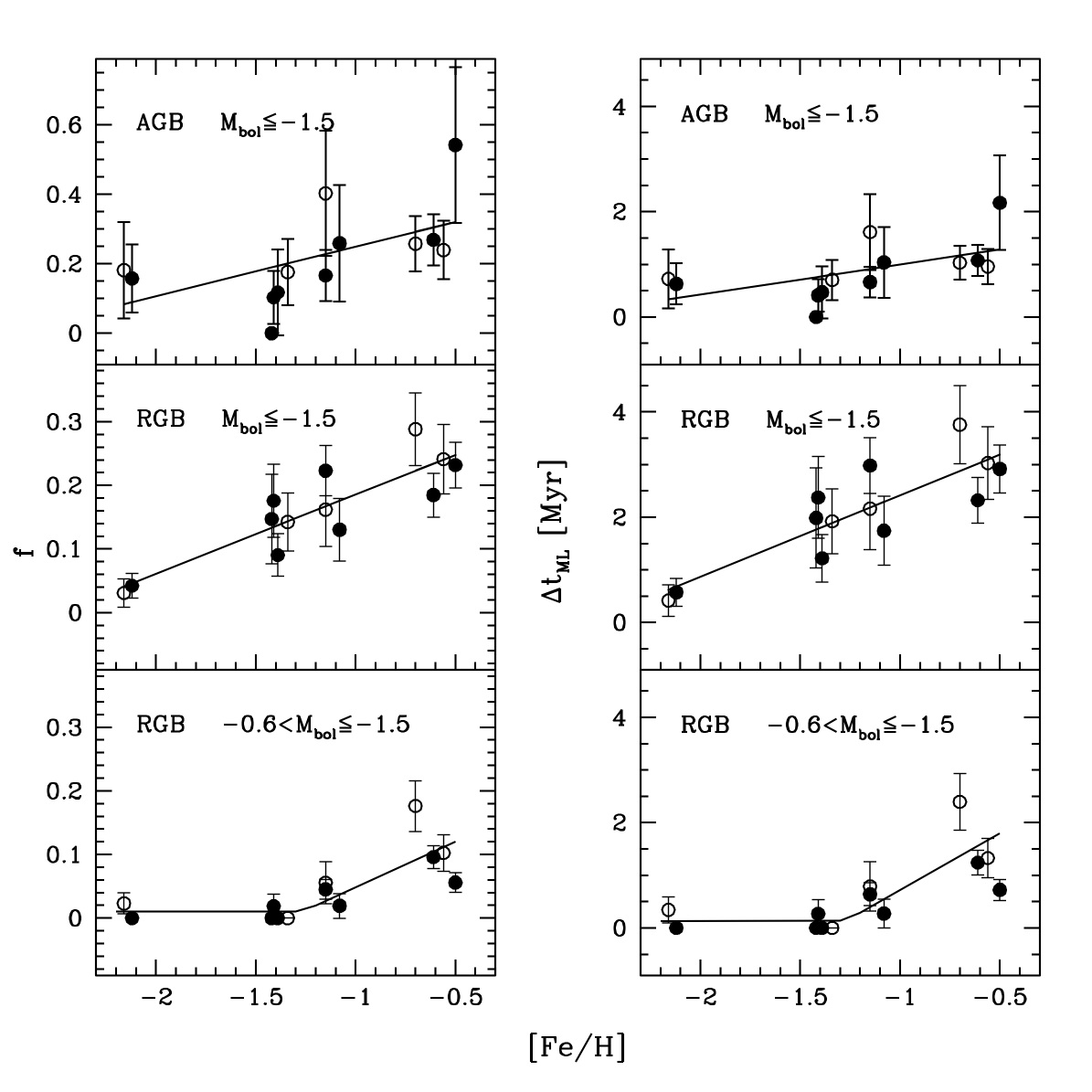}
\end{center}
\caption{Fraction $f$ of dusty RGB and AGB stars (left panels) and effective ML timescales (right panels) 
as a function of metallicity 
and for different magnitude 
bins (see Sect.~\ref{fnum} for details). Open circles refer to clusters with normal HB, filled circles to 
clusters with extended HB. Solid lines are best-fit relations.}
\label{duty}
\end{figure*}

By inspecting the observed CMDs of Figure~\ref{cmd}, one can clearly
see that {\it i}) all GCs, also the most metal-poor ones, have giant
stars with color excess, {\it ii}) in each GC only a fraction of giant
stars show color excess and this fraction quickly decreases with
luminosity
\footnote{There are very few candidate dusty stars in M~30,
NGC~288, and NGC~6539.
In particular, M~30 has a high central concentration but
an intrinsic low luminosity; NGC~288 has a low central concentration
and low intrinsic luminosity; finally, NGC~6539 has a relatively low central concentration. 
For all the three GCs the resulting CMDs and CCDs are poorly populated in the
bright portion of the RGB.
This prevented us from performing any 
quantitative analysis on their population of dusty stars and mass
loss. Hence, these three GCs have not been used to derive ML timescales and rates.}.

This observational evidence cannot be simply ascribed to the somewhat
stochastic formation of dust, since it cannot
explain the observed trend with luminosity in all GCs.  A natural
explanation of the star-to-star variation in IR-excess is that the ML
process is somewhat episodic and operates only a fraction of the
evolutionary time along the RGB and AGB. 
We also find that {\it iii}) dusty giants at lower luminosity 
are preferentially detected in GCs of higher metallicity.
This is at least in part a matter of sensitivity. For a given CS envelope mass
there is less dust to detect at low metallicity.  

Recently, \citet{gro12} also found dust excess in relatively low-luminosity 
Population I giants down to M$_{bol}\approx -1.3$, which is well below the RGB tip.

The fractional
number of giant stars with color excess is an important quantity.
In fact, if the color excess is a signature of a
dusty CS envelope due to ML, it can give an empirical estimate
of the so-called duty cycle -- the fraction of the evolutionary
time along the RGB and AGB during which ML is active \citep[see
Section 4 of][]{ori07}.  
To properly evaluate it 
one needs {\it i}) to separate RGB from AGB stars and 
{\it ii}) to correct observed star counts for possible incompleteness and field contamination.

The ($K,~J-K$) plane is not a very effective means to well separate 
the AGB and RGB evolutionary sequences 
and identify the AGB stars.
Hence, we used complementary high-resolution HST
where available, and ground-based wide field photometry in the
optical bands and we computed 
suitable CMDs in the $V,~(U-V)$, $V,~(B-V)$ and/or
$V,~(V-I)$ planes,  where the RGB and AGB sequences are better separated. 
References to the already published photometry are listed in Table~\ref{phot}.
The photometric reduction of those datasets not published yet was carried
out using DAOPHOTII/ALLSTAR \citep{stet87,stet94}. 
We also used the \citet{cle01} compilation to
identify long period variable stars.

We estimated the completeness of our catalogues by making extensive use 
of artificial star tests.
The degree of completeness of the near-IR photometric catalogues is
always $\approx 100$\% over the full magnitude range covered by our
Spitzer survey.
On the contrary, the completeness of the Spitzer-IRAC photometric catalogues,
which is dominated
by non-dusty giants, is not 100\%. 
It is normally $>80$\% down to $M_{\rm
bol}=-1.5$, with the exception of the very central $10\arcsec$--$20\arcsec$
regions of the most concentrated clusters (NGC~6440, NGC~6441,
NGC~6388) where it can drop down to $\le60$\%\footnote{The cluster centers have been estimated typically
from IR photometric catalogs by adopting the procedure described in  \citet{mon95}.}.
At fainter $M_{\rm
bol}>-1.5$ magnitudes, the degree of completeness depends on both
crowding and distance and does not exceed $\approx$60\% in most
clusters.

To estimate the degree of possible field contamination, we used 2MASS
and selected an annular region at a distance larger than the cluster tidal radii,
typically at $20\arcmin<r<22\arcmin$ around each
cluster and counted the stars within the same color and magnitude range
of the sampled Spitzer population.  Field contamination is negligible
in most clusters except in NGC~6388, NGC~6440, and NGC~6441 where it
turns out to be $\approx3$\%, $\approx19$\% and $\approx20$\% at
$M_{\rm bol}\le -1.5$, and $\approx5$\%, $\approx32$\% and $\approx27$\%
at $-1.5<M_{\rm bol}\le -0.6$, respectively.

We computed the
fractional number $f=n_{dusty}/n_{tot}$ defined as the number of candidate dusty stars divided by the
total number of stars, as counted in the near-IR sample, in suitable luminosity bins,
by correcting star counts for incompleteness and field contamination
and separating RGB from AGB and LPV stars.
To estimate the values of the episodic ML timescale, we
use the fractional numbers $f$  
in each luminosity bin along the RGB and AGB,
and we multiply them for the corresponding evolutionary time
${\Delta t}$. 
The evolutionary times are derived from canonical evolutionary tracks
\citep{pie06} for low-mass (0.8--$0.9\,M_{\odot}$) RGB stars at the
metallicity most suited for each individual GC, $Z$ ranging between
0.008 for the most metal-rich and 0.0003 for the most metal-poor.
The total time ${ \Delta t^{\rm ML}}$ during which a star can lose mass
is thus given by the simple formula
\begin{eqnarray}
{\Delta t^{\rm ML} =  \Delta t \times f.}
\,
\end{eqnarray}

The fractional numbers $f$ of dusty RGB
stars were computed in two suitable luminosity intervals,
namely $M_{\rm bol}\le-1.5$ and $-1.5<M_{\rm bol}\le-0.6$.  These two bins correspond to 
approximately equal evolutionary times of $14\pm1$\,Myr at all
metallicities.  
We did not consider in this analysis fainter RGB stars, since 
only a few metal-rich giants show color excess, 
therefore star counts are
more uncertain because of the intrinsic low number statistics, larger 
corrections for incompleteness, and
lower photometric accuracy near the detection limit.
The fractional numbers of dusty AGB stars
apply only to the brightest luminosity bin at $M_{\rm bol}\le-1.5$,
where most AGB stars have been detected.  This luminosity bin corresponds
to an evolutionary time of $\approx4$\,Myr at all metallicities.

Our final observed numbers $n_{dusty}$ of dusty RGB and AGB stars 
as well as the fractional numbers $f$ (corrected for 
incompleteness and field contamination) in each GC are reported in Table~\ref{phot}. 

In the brightest luminosity bin such a fractional number varies from a few hundredths to a few tenths by 
increasing metallicity, 
while in the lower luminosity bin it can reach about one tenth  
in the most metal-rich clusters.
The typical Poissonian error
$\Delta f/f=\sqrt{(1+f)/n_{dusty}}$ of the estimated fractional number ranges between 30\% and 60\%.
The fractional numbers of dusty AGB stars at $M_{\rm bol}\le -1.5$ 
also increase with metallicity, ranging from
$\approx0.1$ in the metal-poor GCs to $\approx0.3$ in the metal-rich ones with typical errors of 50\%.

There is no significant trend between the fractional number of dusty giants 
and the cluster concentration \citep{har96}.
Indeed, for a given metallicity range, very similar fractional numbers are measured in GCs with 
concentrations ranging from 1.5 to 2.5.

Figure~\ref{duty} (left panels) shows the fractional numbers of dusty RGB and AGB stars in each
GC of our sample for different magnitude bins and  
corresponding ML timescales (with typical uncertainty of $\le 0.5$\,Myr). 
We also included the values for 47~Tuc \citep{ori07}, as computed in the same magnitude bins.
In the following, 47~Tuc has been classified as a metal-rich GC with normal HB.
At $M_{\rm bol}<-1.5$ the
inferred ML timescales (right panels of Figure~\ref{duty}) for RGB stars increase by a factor of
$\approx6$ (from $\approx0.5$ to $\approx3$\,Myr) when increasing
metallicity from ${\rm [Fe/H]}\approx-2.2$ to ${\rm
[Fe/H]}\approx-0.5$, while for AGB stars the ML timescales increase
only by a factor of $\approx3$ (from $\approx0.5$ to
$\approx1.5$\,Myr).  At lower luminosities ML is only detectable
around metal-rich RGB stars on a $\approx1.5$\,Myr timescale.
There are not appreciable differences in the inferred ML timescales 
between clusters with normal and extended HB.

\section{Dust excess and ML rates}
\label{general}  

\begin{figure}[]
\begin{center}
\includegraphics[width=9cm]{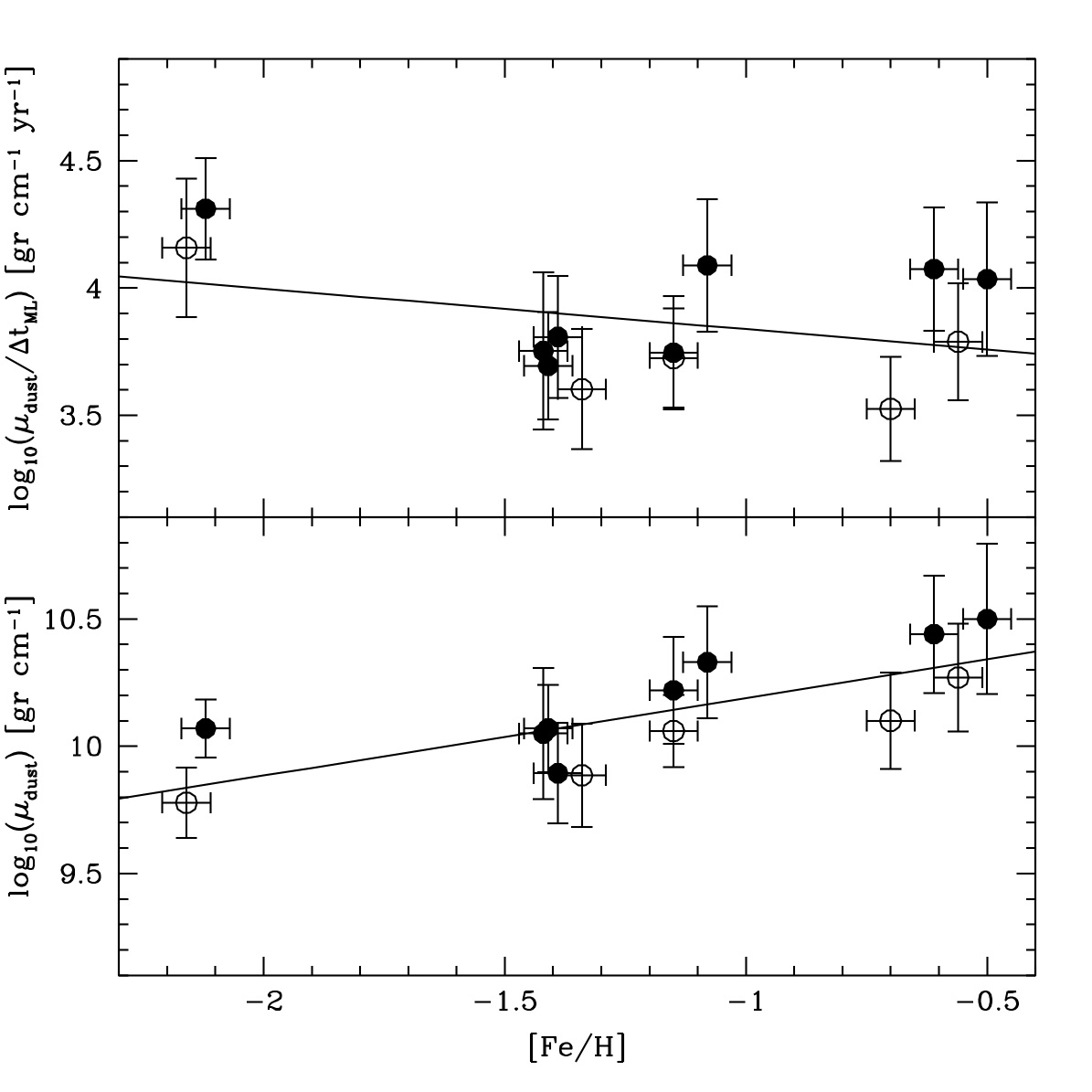}
\end{center}
\caption{
Average $\mu _{\rm dust}$ (bottom panel)
and the $\mu _{\rm dust}$/$\Delta t^{\rm ML}$ ratio (top panel)
for the stars with $M_{\rm bol}<-1.5$ in each GC as a function of the cluster metallicity.
Open circles refer to clusters with normal HB, filled circles to 
clusters with extended HB. Solid lines are best-fit relations.}
\label{dust}
\end{figure}

For the candidate dusty stars in each GC, ML rates can be estimated by
properly modeling the circumstellar dust emission.  To obtain the ML
rates we use our modified version \citep{ori07} of the DUSTY code
\citep{ive99,eli01} to compute the emerging spectrum and dust emission
at the IRAC wavelengths.  We adopt Kurucz model atmospheres for the
heating source and assume the dust to be a mixture of warm silicates
with an average grain radius $a=0.1\,\mu $m.  We tested that slightly different
choices for the dust properties have negligible impact in the
resulting IRAC colors and ML rates.  While radiation pressure acting
on dust might plausibly drive winds in luminous, metal-rich red giants
\citep{wil00}, the GC stars are generally neither luminous nor metal
rich enough for this mechanism to be efficient \citep{ori07,gro12}.  Hence, 
we run the DUSTY code under the general assumption of an expanding
envelope at constant velocity $v_{\rm exp}$ with a density profile
$\eta \propto r^{-2}$, a dust temperature for the inner boundary
$r_{\rm in}$ of 1000\,K and a shell outer boundary $ r_{\rm out}=1000
\times r_{\rm in}$. The average  $r_{\rm in}$ value  in the observed giants turns out to be $\approx 10^{14}$ cm, 
corresponding to tens of stellar radii.
We then computed a large grid of DUSTY models
with stellar temperatures in the 3500--5000\,K range and optical depths
at 8\,$\mu $m ($\tau_8$) between $10^{-5}$ and $10^{-1}$.  For each
candidate star with dust excess, we enter the grid with its empirical
stellar temperature (as derived from its $(J-K)_0$ color, see Sect.~\ref{obs})
and $(K-IRAC)_0$ colors, and we exit the grid with
predictions for the dust optical depth, emerging flux, and envelope
radius.

In the range of temperatures and colors shown by the giants 
sampled in our Spitzer survey, for a given input $(K-8)_0$ color, 
if the input photospheric temperature varies by ~100K, the output opacity $\tau_{8\mu m}$ 
varies by ~10\% and the amount of dust by less than 5\%. 
Hence, the input photospheric temperature is not a major critical parameter in the determination 
of the dust parameters, unless it is uncertain by more than several hundred degrees, 
which is normally never the case for GC giants. 

As discussed in \citet{ori07},
the ML rates are computed by using the formula:
\begin{eqnarray*}
dM/dt =  4 \pi r_{\rm out}^2 \times \rho_{\rm dust} \times v_{\rm exp} \times \delta
\end{eqnarray*}
where $\rho_{\rm dust}\propto\rho_{\rm g} \tau_8 F_8({\rm obs})/F_8({\rm mod}) D^2/r_{\rm out}^2$ 
is the dust density, 
$\rho_{\rm g}=3\,{\rm g\,cm^{-3}}$ is the grain density, 
$F_8$ are the observed and model dust emission at 8\,$\mu$m, 
$D$ the GC distance, $r_{\rm out}$ the envelope outer radius and $\delta$ the gas-to-dust ratio.

The $v_{\rm exp}$ and $\delta$ quantities and especially their scaling with metallicity 
are somewhat free parameters. 
At variance, the linear density of 
dust mass $\mu _{\rm dust}=4 \pi r_{\rm out}^2 \times \rho_{\rm dust}$
depends on observed quantities.  Its absolute value also depends on
the assumed individual grain density and overall circumstellar geometry, 
which, however, are normally considered 
independent of the metallicity of the central star.
This is also the only working scenario that one can approach with the today's knowledge of the field, 
given that we do not know whether and how dust and circumstellar envelope properties, especially 
around low-mass giants, may or may not change with metallicity.

\subsection{Dust ML}
\label{dustML}

Figure~\ref{dust} (bottom panel) shows the average $\mu _{\rm dust}$  
for the stars with $M_{\rm bol}<-1.5$ in each GC of our sample as a function of the cluster metallicity.
We also included the value for 47~Tuc \citep{ori07}, as computed in the same magnitude bin.
The linear dust mass density increases by a factor of $\approx$3 by increasing metallicity from [Fe/H]=-2.2 
to [Fe/H]=-0.5.
By computing the ratio between $\mu _{\rm dust}$ and $\Delta t^{\rm ML}$ 
one can also get an estimate of the linear dust mass density rate and its behavior 
with varying metallicity (see Figure~\ref{dust}, top panel).
Such a rate decreases by a factor of $\approx$2 with increasing [Fe/H] from -2.2 to -0.5. 
 
We derived these trends without any specific assumption on the metallicity
dependence, hence they can be considered as an observational finding. 

\subsection{Gas + dust ML rates: assumptions}
\label{total}

If dust and gas are coupled and no mechanism preferentially removing one of the two 
components is at work, 
one should expect to get similar trends 
for total (gas + dust) ML rates and amount of mass lost with varying metallicity 
as observed for the dust alone.
Hence, we can now attempt to compute total ML rates under {\it reasonable} 
assumptions for the gas-to-dust ratio and the expansion velocity of the envelope.

A lower limit to $\delta$ is given by $1/Z$ \citep[see, e.g.,][]{ler07}, 
where $ Z=(10^{{\rm [M/H]}}\times Z_{\odot}$) is the global
metallicity (see Table~\ref{clusters}).   
A slower than 1/Z scaling for the gas-to-dust ratio would imply a 
concentration of metals higher in the envelope than in the photosphere, 
which is problematic unless for some reasons a large fraction of gas (hydrogen) 
is quickly escaping from the envelope while dust does not. 
At variance, a faster scaling is always possible but it will quickly raise the ML rates 
to prohibitive values at low metallicity, resulting in a total ML exceeding the stellar mass, 
which is unphysical.

As a reference zero point value for the gas-to-dust ratio we use 
$\delta_0\approx 200$ at the global metallicity ($Z_0\approx 0.005$) of 47~Tuc, so  
$\delta=\delta_0\times Z_0/Z$.

This zero point value is also a somewhat lower limit, 
requiring that already $\approx$50\% of the $\alpha$-elements 
(O and Si in particular, among the main constituents of dust grains in these low-mass stars) 
available from the star atmosphere 
(and typically enhanced by a factor of 2-3 with respect to the iron abundance in GCs) 
condensate into dust once reaching the equilibrium radius at a distance of several tens/hundreds stellar radii. 
On the other hand, such a zero point value cannot be much higher 
(or equivalently the condensation fraction much lower than 50\%), otherwise the total mass lost will quickly 
exceed the stellar mass. 

In summary, both a faster than 1/Z scaling of the gas-to-dust ratio and/or a 
condensation fraction significantly 
lower than 50\% would result in unphysically high ML. 
On the other hand, a slower than 1/Z scaling of the gas-to-dust ratio 
would result in an anomalously high (even higher than in the photosphere) concentration of metals 
in the envelopes of metal-poor stars. 
The condensation fraction of metals into dust grains cannot be much higher (a factor of 2 at most) 
than 50\%, by definition.

Given that low-mass giants 
have very similar gravities, hence very similar escape velocities, regardless of their metallicity, 
for a given input energy the expansion velocity $v_{\rm exp}$ of their circumstellar envelopes 
is expected to scale as $\delta ^{-0.5}$ if dust and gas are coupled. 
Indeed, by increasing the number of gas particles (i.e., for higher
value of $\delta$), the momentum per particle (either gas or dust) is
smaller, hence $v_{\rm exp}$ decreases \citep{hab94,van00}, independent of the nature 
of the input energy source.  
The alternative assumption of both a constant or increasing expansion velocity with decreasing metallicity 
would require a much higher input energy in metal-poor stars, 
which is possible, given also that such an energy source is unknown,  
but it would also quickly raise their ML to unphysically high values, 
exceeding the stellar mass.

As a reference zero point value for the expansion velocity, we adopt $v_{\rm exp}^0=10~{\rm
km\,s^{-1}}$ at the metallicity/gas-to-dust ratio of 47~Tuc, and we
scale it accordingly to $(\delta_0/\delta)^{0.5}$.  This gives values
similar to typical expansion velocities measured in luminous, nearby
giants \citep[see, e.g.,][]{net93,so01}, which can range between a few and
$\approx 20\,{\rm km\,s^{-1}}$.  
Very recently, \citet{gro14} detected rotational CO line emission in a nearby,
low luminosity giant and measured an expansion velocity of 12 km/s.
Even in the most metal-poor stars
where $v_{\rm exp}\approx 1\,{\rm km\,s^{-1}}$, $v_{\rm exp}$ exceeds
the sound speed.
These envelope velocities are typically lower than the wind velocities derived from the measurements of 
chromospheric lines in field giants \citep[see, e.g.,][]{dup09}, although such lines are better tracers of the region where 
wind forms and accelerates. 

\subsection{Gas + dust ML rates: results}
\label{results}

\begin{figure}[htbp]
\begin{center}
\includegraphics[width=9cm]{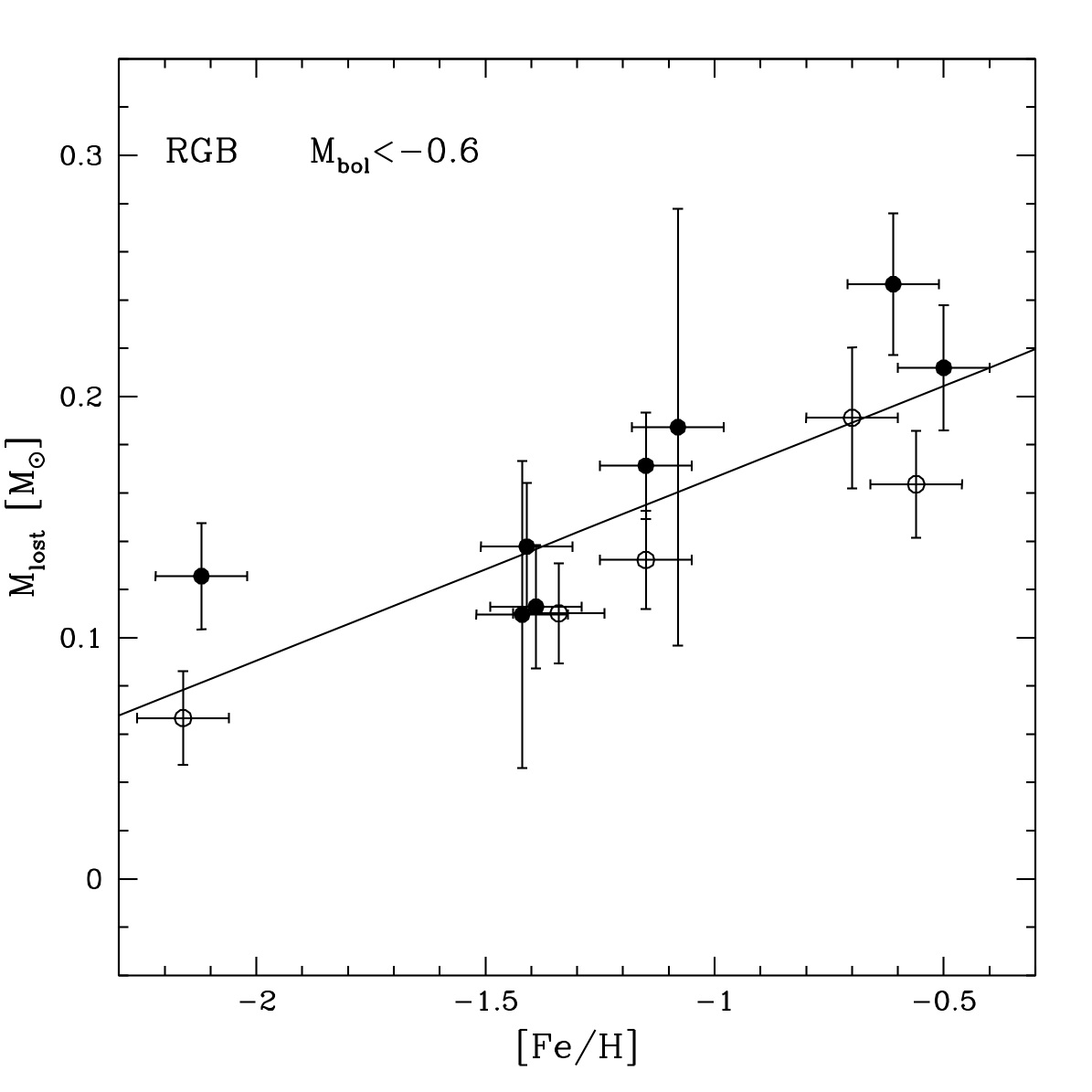}
\end{center}
\caption{
Total ML on RGB  as a function of metallicity. For each GC, rates are computed by averaging
the values of the single stars, duty cycles from the fitting relation. Only the two
upper magnitude bins are used.
Empty circles: GCs with normal HB. Filled circles: GCs with extended HB.
Solid line: best-fit relation.}
\label{mtotrgb}
\end{figure}

\begin{figure}[htbp]
\begin{center}
\includegraphics[width=9cm]{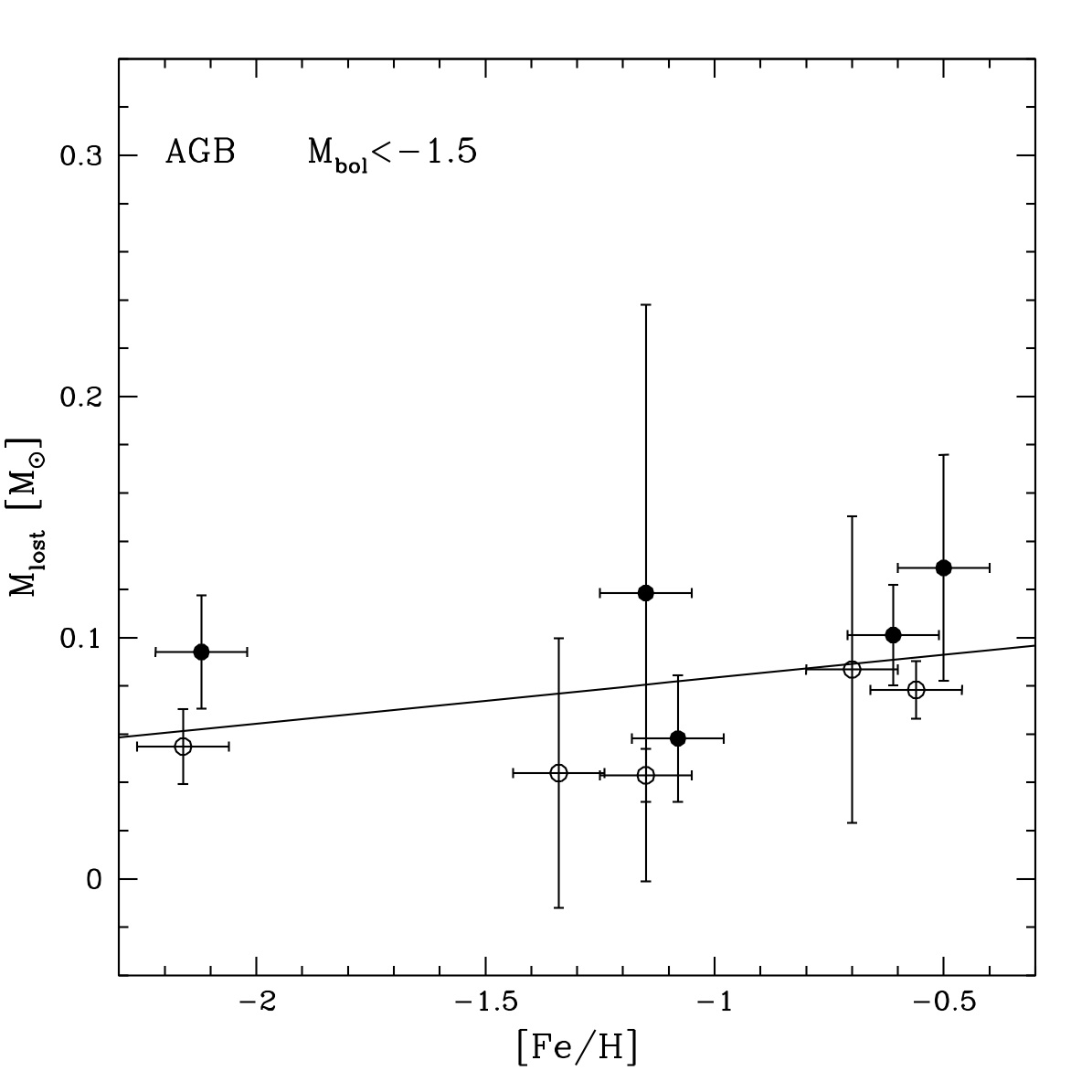}
\end{center}
\caption{
Total ML on AGB  as a function of metallicity. For each GC, rates are computed by averaging
the values of the single stars, duty cycles from the fitting relation. Only the
upper magnitude bin is used.
Empty circles: GCs with normal HB. Filled circles: GCs with extended HB.
Solid line: best-fit relation.}
\label{mtotagb}
\end{figure}

Under the assumptions for the zero point values and trends with metallicity of the gas-to-dust ratio and expansion 
velocity discussed in Sect.~\ref{total},
we can compute ML rates for RGB and AGB stars as
a function of luminosity normalized using the Reimers parameter,
$(L_*/g_*R_*)_{\odot}$, where $L_*$, $g_*$, and $R_*$ are the stellar
luminosity, gravity, and radius in solar units\footnote{The Reimers
parameter is a simple combination of stellar parameters with units of
ML.}.

Our best-fit empirical formulae for such ML rates in units of [\msun\,{\rm yr}$^{-1}$] are as follows.
\begin{eqnarray*}
{(dM/dt)_{\rm RGB} = C \times 4.57 \times 10^{-10} \times (L_*/g_*R_*)_{\odot}^{0.43} \times f({\rm [Fe/H]})},
\,
\end{eqnarray*}
\begin{eqnarray*}
{(dM/dt)_{\rm AGB} = C \times 2.24 \times 10^{-11} \times (L_*/g_*R_*)_{\odot}^{0.70} \times f({\rm [Fe/H]})},
\,
\end{eqnarray*}
where
\begin{eqnarray*}
{C = (\delta_0/200)^{0.5}\times (v_{\rm exp}^0/10)\times(\rho_{\rm g}/3)},
\end{eqnarray*} 
and 
\begin{eqnarray*}
{f({\rm [Fe/H]}) = 10^{-0.25\times ({\rm [Fe/H]}+0.7)}}.
\end{eqnarray*} 
The $rms$ uncertainty of these ML rate fitting formulae is $\approx$30\%.

The ML rates increase with increasing luminosity and with decreasing
metallicity.  Our best-fit relations for RGB and AGB stars have
shallower slopes with varying luminosity than the Reimers' law
\begin{eqnarray*}
{dM/dt} =  \eta_{\rm R} \times \, 4 \times 10^{-13} \times (L_*/g_*R_*)_{\odot} \qquad [\msun\,{\rm yr}^{-1}].
\end{eqnarray*}

The latter, however, can still fit our AGB ML rates at $\approx 2\sigma$
level of confidence.

The ML rates can be also conveniently computed as a function of bolometric
magnitude and metallicity.  
Our best-fit formulae are as follows.
\begin{eqnarray}
{\log_{10} (dM/dt)_{\rm RGB} = C' - 0.28 \times M_{\rm bol} - f'({\rm [Fe/H]}) - 8.00},
\end{eqnarray}
\begin{eqnarray}
{\log_{10} (dM/dt)_{\rm AGB} = C' - 0.40 \times M_{\rm bol} - f'({\rm [Fe/H]}) - 8.25},
\end{eqnarray}
where
\begin{eqnarray*}
{C' =  \log_{10} C }
\,
\end{eqnarray*}
and%
\begin{eqnarray*}
{f'({\rm [Fe/H]}) = 0.25\times ({\rm [Fe/H]}+0.7)}.
\,
\end{eqnarray*}

According to the inferred best-fit relations, ML rates (both in AGB and RGB) 
decrease by a factor of $\approx$2.6 with increasing [Fe/H] from -2.2 to -0.5. 
This factor is fully consistent with the factor of $\approx$2 obtained for the average linear mass density rates 
of the dust alone (mostly an observational finding). 

Such a similarity of trends seems to suggest that 
(i)~the assumed scaling laws for the gas-to-dust ratio and expansion velocity with metallicity,
based on common-sense/current-knowledge physics, 
are self-consistent with the assumption of gas and dust being coupled in the outflow   
of low-mass giants, and (ii)~ the modest anti-correlation between ML rates and metallicity is 
not a mere consequence of assumptions. 

\section{Total Mass Loss}
\label{totalml}
Once rates and timescales at different stellar luminosity 
and metallicity are estimated, total ML at a given metallicity can be computed by integrating over the 
evolutionary time along the RGB and AGB.

By using the simple equation: 
\begin{eqnarray*}
\Delta {M}  = \Sigma_i( {dM/dt}_i \times \Delta {t_i^{\rm ML}} ) 
\end{eqnarray*}
and multiplying the average ML rate by the ML timescale (see eq. 1) 
in each {\it i-th} luminosity bin, we find the total ML on the RGB and AGB.
For each GC, including 47~Tuc, average ML rates are computed by averaging
the values obtained for the individual stars. 

Figure~\ref{mtotrgb} shows total ML in RGB as a function of metallicity,
summing the contribution of the two magnitude bins defined before, 
i.e. at
$M_{\rm bol}<-0.6\,M_{\odot}$. 
In a similar
fashion, Figure~\ref{mtotagb} shows total ML in AGB as a function of
metallicity at $M_{\rm bol}<-1.5\,M_{\odot}$.

The resulting fitting formulae are:

\begin{eqnarray}
{ML^{\rm RGB} =  0.08 \times {\rm [Fe/H]} + 0.24 \pm 0.03~(rms)}
\qquad [\msun],
\end{eqnarray}

\begin{eqnarray}
{ML^{\rm AGB} =  0.02 \times {\rm [Fe/H]} + 0.10  \pm 0.03~(rms)}
\qquad [\msun]. 
\end{eqnarray}

If we separately fit clusters with normal and extended HB,
we find fitting formulae with the same slope and minor ($\mp0.02$) zero point variations.

If we compute the average ML rates by means of eqs. 2,~3 instead of averaging 
the values of individual stars, we still obtain very similar fitting formulae 
within 1$\sigma$ uncertainty.

It is interesting to note that 
the total amount of lost gas and dust increases by a factor of $\approx$3 with increasing [Fe/H] from -2.2 to -0.5 
as in the case of the total amount of lost dust alone (see Sect.\ref{dustML}). 

The inferred absolute values for the total ML are somewhat lower limits.
Indeed, as mentioned in Sect.~\ref{general}, we adopt a lower limit for
the gas-to-dust ratio $\delta \propto1/Z$, and ML rates are
proportional to $\rm \sqrt{\delta}$.  Hence, if for example, we assume
that $\delta \propto 1/(Z/2)$, the total ML
would increase by $\sqrt{2}$, i.e., by $\approx40$\%.  Moreover,
although the bulk of the ML occurs in the brightest portion of the RGB,
some ML can also occur at lower luminosities than those sampled by our
Spitzer survey.

\section{Discussion and conclusions}

We have inspected the near- and mid-infrared color-magnitude and color-color diagrams of  
a carefully
chosen sample of 15 Galactic GCs spanning the entire metallicity
range from about one hundredth up to almost solar and, for a given metallicity, with different
HB morphology.

All GCs, including the most metal-poor ones, have RGB and AGB giant
stars with color excess, plausibly due to dust formation in mass flowing from them. 
Such dusty giants are detected down to M$_{bol}\le -1.5$ at all metallicities and 
down to M$_{bol}\approx$0 in the most metal-rich GCs.

We find that the fractional number of giants stars with color excess 
increases towards higher luminosities and metallicities.

By modeling the mid-infrared color excess of our sample of GC giants,
we are able to derive ML rates in a representative sample of
Population II RGB and AGB stars with varying metallicity. At a given
$M_{\rm bol}$ only a fraction of stars are losing mass\footnote{This evidence is in agreement with the
consideration that it is impossible for
a low-mass ($\approx 0.8\, M_{\odot}$) giant to lose mass at the
estimated rates (see Sect.~\ref{results}) during the entire time of
its ascent of the RGB and AGB, simply because it would eject an amount
of gas exceeding its total mass.}. From this, we
conclude that the ML is episodic. The observed fraction of dusty
giants gives the time that the ML is ``on.'' Combining this duty
cycle with the ML rates yields the total ML.  In the following
subsections, we summarize our findings about ML and its possible dependence on the
metallicity and HB morphology of the parent cluster.

\subsection{Mass loss and metallicity}

Our estimates of ML in Population II RGB stars indicate that ML
depends only moderately on metallicity.  Indeed, ML rates slowly decrease
with increasing metallicity, while duty cycles more rapidly increase
with increasing metallicity, with the net result that total ML
moderately increases with increasing metallicity, about 0.08\,\msun\
every dex in [Fe/H].  By using an indirect method based on the
estimate of stellar masses on the HB, \citet{gra10} find a similar
dependence of total ML on metallicity.

The ML rates in Population II AGB stars show a similar dependence on
metallicity as RGB stars, while duty cycles increase more slowly with
it (see Sect.~\ref{fnum}).  We estimate $\le 0.1$\,\msun\ of total ML
on the AGB, nearly constant with varying metallicity.

The fact that ML rates in both Population II RGB and AGB stars
seem to increase with decreasing metallicity, although rather slowly, 
would suggest that the outflow cannot be mainly driven by mechanisms involving
opacity from metals.

\begin{figure}[]
\begin{center}
\includegraphics[width=9cm]{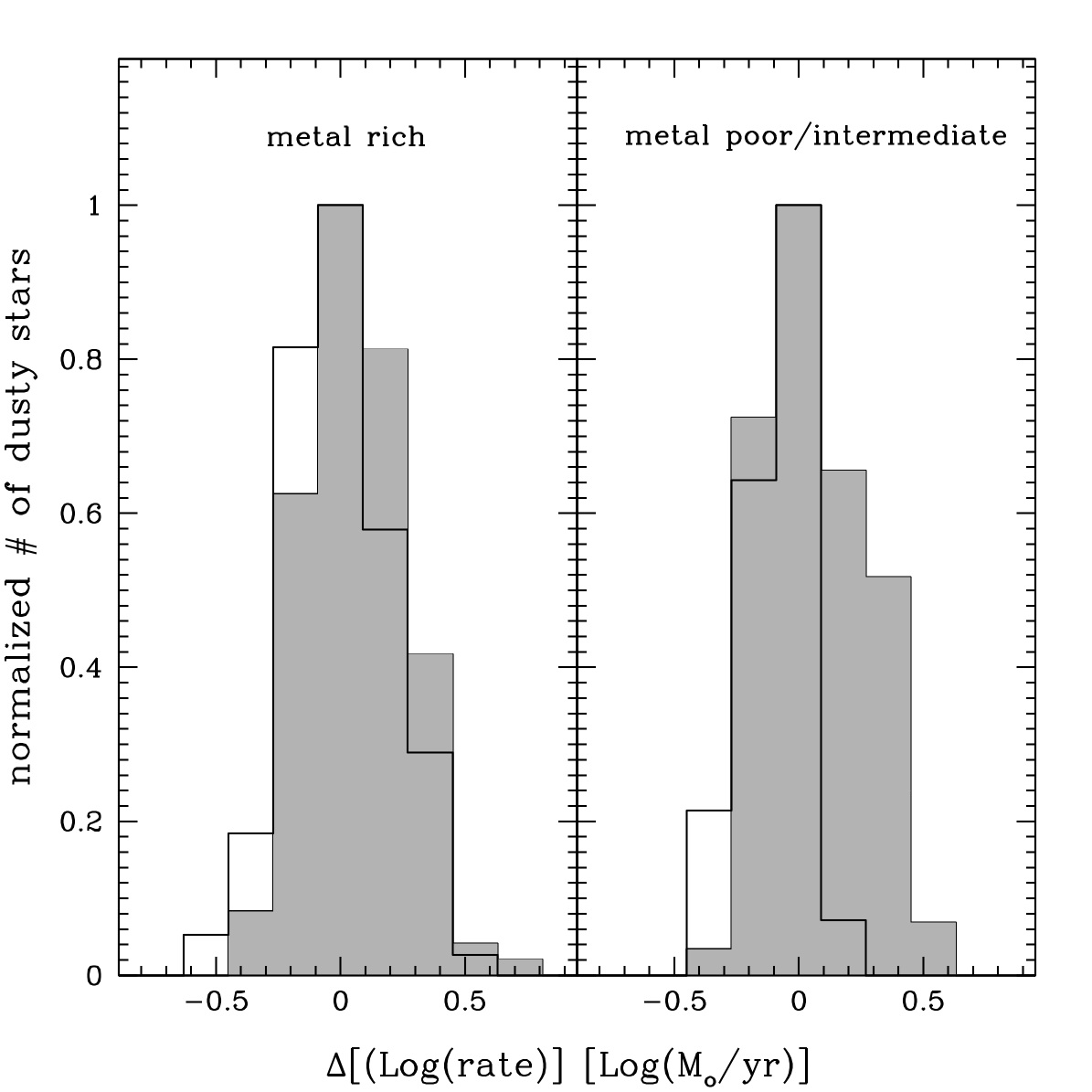}
\end{center}
\caption{Global histograms of $\Delta$[Log(ML rates)] (${\rm
measured-best fit}$) for metal-rich (left panel) and metal
intermediate/poor (right panel) GCs, grouped in two subsamples,
namely with normal (empty histograms) and extended (gray histograms)
HB.  }
\label{isto}
\end{figure}

\begin{figure}[]
\begin{center}
\includegraphics[width=9cm]{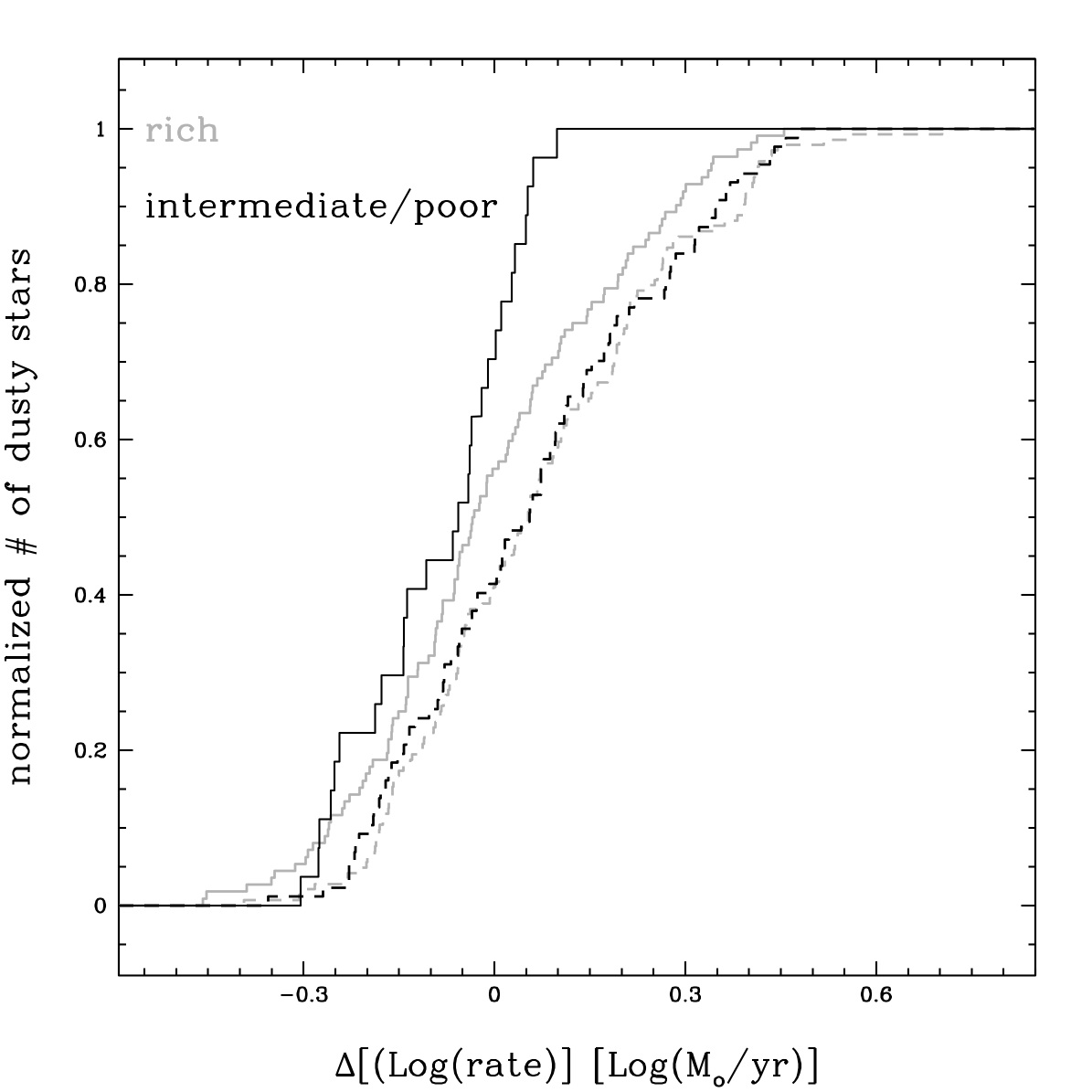}
\end{center}
\caption{Cumulative distributions of $\Delta$[log(ML rates) (${\rm
measured-best fit}$) for metal-rich (gray lines) and metal
intermediate/poor (black lines) GCs, grouped in two subsamples,
namely with normal (solid lines) and extended (dashed lines) HB.}
\label{cum}
\end{figure}

\subsection{Mass loss rates and HB morphology}

The last generation of HST color-magnitude diagrams in the optical
\citep[see, e.g.,][]{ric97,dot10} and UV 
\citep[see, e.g.,][]{fer98,dal13a,dal13b} prove
that the HB morphology of GCs is even more complex than previously
believed and several 2nd parameters can be invoked 
\citep[see, e.g.,][]{roo73,ffp75,ffp76,ren77,ffp93,cru96,cat09,dot10,gra10}.  
Some quantitative investigations of the HB morphology of the massive GCs 
NGC~2808, NGC~6388, and
NGC~6441 HB, were recently performed
\citep{bus07,dal08,bro10,dal11}.

A significant population
of blue, extreme blue, and blue hook HB stars (hereafter BHB, EHB and
BHk, respectively) was found.  
For example, in the metal-intermediate GC
NGC~2808 \citet{dal11} account for 39\% BHB, 11\% EHB, and 9\% BHk, while in
the metal-rich GC NGC~6388 \citet{dal08} account for 15\% BHB, 2\% EHB, and
2\% BHk.  In these GCs, HB models with normal He abundance
($Y\approx0.24$) and ML can account for red HB stars. 
On the contrary the hotter BHB and EHB
could be explained by a higher He content.  BHk stars are
extremely hot HB stars with a significant spread in luminosity, likely
due to a delayed, hot He-flash.  It has been suggested that these stars
could have experienced an enhanced ML during the RGB evolution
\citep{cas09,moe07,dal11}, or alternatively, they could have an
extremely large He content ($Y>0.5$) \citep{dan08} due to extra mixing processes 
undergone during their RGB phase.  

In the following, we briefly 
explore whether and in which terms our 
results on ML could eventually provide additional constraints to 
these working scenarios. 

We computed the ratio between the measured ML rates in each RGB star
and the corresponding best-fit value or equivalently the difference of
their logarithmic values. We then constructed global histograms of
$\Delta$log(ML rates) for metal-rich ([Fe/H]$>-1.0$) and metal-intermediate/poor ([Fe/H]$\le-1.0$)
clusters, grouped in two sub-samples, namely those with normal and
extended HB, as shown in Figure~\ref{isto}.  The histograms of GCs
with normal HBs have Gaussian dispersion $\sigma \approx 0.2$ dex
(rich) and $\sigma \approx 0.12$ dex (metal-intermediate/poor).  A
larger dispersion of $\Delta$log(ML rates) in metal-rich GCs is not
surprising, given that these stars have a larger {\it turnoff} mass and a
wider range of possible masses in the red part of the HB.  The
histograms of GCs with extended HBs have Gaussian dispersion similar
to those of GCs with normal HBs, but have a tail (that is an excess of
stars) toward higher ML rates.

Independent of metallicity, the bulk ($>90$\%) of RGB stars in GCs with
normal HB have rates within a factor of two from the average value.
In GCs with extended HB about 15\% of RGB stars have ML rates in
excess by a factor of two (i.e., by 2--3$\sigma$) from the average
value.

We also computed cumulative distributions $\Delta$[log(ML rates)] for the ML rates, as shown
in Figure~\ref{cum}.  The cumulative distribution of $\Delta$[log(ML
rates)] in metal-rich GCs with normal HB is more bent than the
corresponding distribution for metal-intermediate/poor GCs, in
agreement with the larger Gaussian dispersion.  The cumulative
distributions of $\Delta$[log(ML rates)] in GCs with extended HB are
also more bent (particularly for metal-poor GCs) and shifted toward
higher ML rates, compared to those GCs with normal HBs.
The KS-tests give probabilities of $\approx 5$\% (metal-rich) and $<0.1$\%
(metal-intermediate/poor) that normal and extended HB
distributions be extracted from the same parent population.

By comparing the $\approx 15$\% estimated percentage of stars with ML
rates in excess by a factor of two from the average values with the HB
population ratios in NGC6388 (metal-rich) and NGC2808 (metal
intermediate) reported above, we can speculate that: (1) metal-rich GC
RGB stars with ML rates within a factor of two from the average value
will probably evolve as red clump stars or moderate BHB, depending on
their actual ML rate, duty cycle, and He content,
while those with the
highest ML rates will likely evolve as hot BHB, EHB, or BHs stars.  It
is also possible that those stars with extreme ML rates will move
directly to the WD cooling sequence, without experiencing any
He-flash; (2) metal-intermediate/poor GC RGB stars with ML rates
within a factor of two from the average value can evolve either as red
or BHB and EHB stars, depending on their actual ML rate, duty cycle,
and He content.  Those RGB stars with the highest ML rates (in excess
by a factor of two from the average value) can be precursors of the
hottest EHB and BHk stars.

In practice, for a given temperature on the HB, there can be a certain level
of degeneracy between ML and He content, the two parameters being
somehow anti-correlated.  Indeed, according to evolutionary tracks
\citep{pie06} with normal and enhanced He content, for equal age and
metallicity, a star with higher He content has a smaller {\it
Turnoff} mass compared to a star with normal He, hence the former
should need less ML than the latter to reach a given temperature on the HB.

\begin{acknowledgements}
The authors dedicate this paper to the memory of Bob Rood, a pioneer scientist in the theory of low-mass star evolution,
who inspired this work,  and a friend, who sadly passed away on November 2nd, 2011.
This research is part of the COSMIC-LAB (http://www.cosmic-lab.eu/Cosmic-Lab/Home.html) project funded 
by the European Research Council (under contract ERC-2010-AdG-267675). 
\end{acknowledgements}

\newpage

\noindent{\bf APPENDIX: The impact of crowding and photometric errors.}\\
In this section, we 
briefly discuss the major biases and sources of errors in our photometric analysis that could potentially  
affect the results.\\

\noindent{\bf A1.  Crowding and blending}\\
Since the Spitzer-IRAC pixel size is relatively large ($\sim1.2\arcsec pixel^{-1}$), it is possible that
more than one star actually falls in it. 
Hence blending due to
crowding is an obvious worry near cluster centers. 
However,
as discussed in \citet{ori10}, our approach of combining the relatively low-resolution IRAC photometry
with high-resolution near-IR photometry (in some cases also supported by HST photometry) has been designed
to minimize this problem. 

In fact, the most common cases of two relatively bright giants falling within the
8\micron PSF, thus potentially mimicking spurious dusty stars, should be easily identified in the higher
resolution near-IR images.
Hence, to avoid any spurious detection of color excess due to blend, 
for each candidate dusty star in the surveyed GCs, 
we directly inspected a $5\arcsec \times 5$\arcsec ~high-resolution and deep $K$ band sub-image
centered on it. 
If we identified star(s) 
within the PSF and with comparable brightness (well within an order of magnitude, 
the exact value depending on their distance from the target)  
at 8\,\micron as well as in the $K$ band 
(given that stars with pure photospheric emission have $(K-8)_0\approx0$),
the target star was rejected as a dusty candidate and not included in the final samples of dusty giants shown in 
Figure~\ref{cmd} and Figure~\ref{ccd}.
In addition, in a few suspect cases we performed the same procedure using HST images in the F814W band, as
already done in \citet{ori10}. This provides the most solid evidence that the IR excess is not due to
blends.\\
As stated in Sect. 4, we performed artificial star experiments to estimate the degree of completeness 
of our Spitzer and near-IR catalogues. 
These experiments have also been used to evaluate the fraction of expected blending due to crowding 
(either cluster or field stars), from a statistical approach. 
We found that this fraction is always below 5\%, 
and  in agreement with the values estimated by the direct inspection of high-resolution near-IR and HST images. \\

\noindent{\bf A2. Unresolved background.}\\
We emphasize that the main source of background noise in the Spitzer images of our GCs
is neither zodiacal light (also tabulated in the header of the fits images) nor unresolved galaxy emissions, 
but, as a matter of fact, unresolved stellar light 
(a few to several times the zodiacal light at 8\micron and fully dominant at shorter wavelengths). 
In the observed clusters, zodiacal light at 8\micron ranges between 4 and 8 el~s$^{-1}$ 
(i.e., between 5 and 10 MJy~sr$^{-1}$, in perfect agreement with the very recent 
estimates by \citet{kri12})\footnote{The output product of the Spitzer 
pipeline are images with signal per pixel in unit of MJy~sr$^{-1}$. 
These values can be easily converted in el~s$^{-1}$ using the conversion factors tabulated in the 
header of each fits image.}, 
while the unresolved stellar background ranges between 10 and 40 el~s$^{-1}$. 
These background levels correspond to Vega magnitudes ranging between 11.5 and 13.5 at 8\micron. 
The large majority of the stars sampled in our survey
are still brighter than the background, especially those we target as candidate dusty stars, 
and only the lower RGB sequence in the most distant clusters is fainter than the unresolved background. 
However, even these fainter stars are always many times brighter than the background noise. 
The PSF fitting
procedure we used (see Section~2) also provides a local estimate of the background
level, hence possible contamination and local variations by diffuse light 
is accounted for and automatically subtracted from the
computed instrumental magnitudes.

\noindent{\bf A3. Photometric errors.}\\
To properly quantify the photometric errors, we define the S/N ratio as the ratio between 
the signal of the star in el~s$^{-1}$ (background subtracted) and the square root of the total (star+background) 
signal in el~s$^{-1}$, multiplied by the square root of the on-source integration time (in sec).
In all clusters, the faintest stars that we measured always have S/N$>$15 and those with color excess 
always have S/N$>$100 in all Spitzer bands. 
This implies that pure dust emission (which is always $>$~30\% of the total light at 8\,\micron)
has always been detected at S/N$>$30 and we can thus firmly exclude that the 8\,\micron flux 
in excess of the photospheric emission is due to background fluctuations. 
In summary, we obtain that random photometric errors of the stars we reliably 
measured are always less than 10\% (i.e., $<$0.1 mag) in all Spitzer bands. 
Such relatively small random errors are not surprising, 
given that the final on-source integration time in each filter of our 
Spitzer observations has been quite long, ranging between 1000 and 2700 sec. 
Our complementary near-IR photometry also has very small (typically $<$0.03 mag) random errors 
(see Valenti et al. 2004, 2004b, 2007 for details).
Zero points calibrations in both the near-IR and Spitzer bands are uncertain by a few percent.

\noindent{\bf A4. 47~Tuc: a test-bench target.}\\
An optimized photometric reduction and overall random errors well below $<$0.1 mag both in the near- and mid-IR bands 
are mandatory in order to safely detect small color excesses. A clear example of how large photometric 
uncertainties in both  pass-bands can lead to misleading results is offered by the case of  47~Tuc. 
Indeed, our finding of dusty stars down to about the HB level \citet{ori07,ori10} has been questioned by 
\citet{mcd11} and \citet{mom12}.
However, an accurate comparison of the datasets has shown that their claims were mostly a consequence 
of an insufficient accuracy of their near- and mid-IR photometry.  
 
In fact, the photometry by \citet{mcd11} has larger photometric errors at any given magnitude than
that presented in \citet{ori10}.
Indeed, by comparing the CMDs in Figs. 1 and 2 of \citet{ori10} with the CMDs in Fig. 12 of \citet{mcd11},  
one can 
clearly see that the color scatter in the latter is significantly larger than in the former.
Moreover, \citet{mcd11} made an extensive use 
of 2MASS photometry in the inner region, which is less accurate and likely     
affected by blending because of its lower resolution than the data used by \citet{ori07}.
In fact, by comparing the K-band and 8\micron photometry for a subsample of candidate dusty 
stars in common between \citet{ori07} and \citet[][see their Fig. 13]{mcd11},
one finds that  8 \,\micron photometries differ by about +0.1 magnitude only, 
while K-band photometries differ by about -0.3 magnitudes, 
implying average 0.2 magnitude bluer (K-8)$_0$ colors for the dusty stars in \citet{mcd11}. 
This indicates that the main discriminant between the two analyses is the K band, not the 
Spitzer photometry. The bluer K-8 colors and the
overall larger errors of \citet{mcd11} photometry largely prevent the detection of color excesses 
(a few tenths of a magnitude) as measured by \citet{ori07}.

Similar arguments apply to the \citet{mom12} photometric survey 
of 47~Tuc using the mid-IR imager VISIR at the VLT. 
While VISIR has an optimal, high spatial resolution, unfortunately its sensitivity is insufficient 
to probe small color excesses along the RGB in GCs.  
Indeed, Fig. 3 and Table 2 of \citet{mom12}  show that 
about two magnitudes below the tip 
their photometric error rapidly increases 
from 0.1 up to 0.3 magnitudes. Such errors are comparable with (at 1-2$\sigma$ level) if not exceeding, the majority 
of color excesses measured by \citet{ori07}. 
This is not surprising,  in fact even at an 8m--class telescope and with the use of narrow band filters, the thermal 
background is still so high and variable that it is very difficult to obtain accurate, 
deep photometry from the ground 
(as required for this scientific goal) compared to a space facility like Spitzer. 

\end{document}